\documentclass[pra,twocolumn,showpacs,floatfix]{revtex4}
\usepackage{graphicx}
\begin{document}

\title{Mixtures of Bose gases confined in concentrically coupled
annular traps}
\author{F. Malet$^1$, G. M. Kavoulakis$^2$, and S. M. Reimann$^1$}
\affiliation{$^1$Mathematical Physics, Lund Institute of Technology,
P.O. Box 118, SE-22100 Lund, Sweden \\
$^2$Technological Education Institute of Crete, P.O.
Box 1939, GR-71004, Heraklion, Greece}
\date{\today}

\begin{abstract}
A two-component Bose-Einstein condensate confined in
an axially-symmetric potential with two local minima,
resembling two concentric annular traps, is investigated.
The system shows a number of phase transitions that
result from the competition between phase coexistence, and
radial/azimuthal phase separation. The ground-state phase diagram,
as well as the rotational properties, including the (meta)stability
of currents in this system, are analysed.
\end{abstract}


\maketitle

\section{Introduction}

The field of cold atoms has expanded dramatically over the
last 15 years. It has now reached a stage where experimentalists
are capable of designing the form of the confining potential.
Going to extreme aspect ratios, conditions of quasi-one- and 
quasi-two-dimensional behavior have been achieved. In other 
experiments, it has also become possible to design toroidal 
trapping potentials \cite{Gupta,Olson,Ryu,Hen,vK}, in which persistent 
currents have been observed \cite{Ryu}.

Recent theoretical studies have examined Bose-Einstein condensates in
one-dimensional annular traps. For example, quantum-tunneling-related
effects in vertically \cite{Wolf} and concentrically \cite{Brand}
coupled double-ring traps were investigated. Also, the rotational
properties of a mixture of two distinguishable Bose gases that are
confined in a single ring were addressed \cite{Smy09}. One of the
basic points of the above studies is the fact that the ability to
design traps, control and manipulate the atoms with a very high
accuracy, may allow the investigation of novel quantum phenomena,
like quantum phase transitions, for example.

In the present work we consider a mixture of two distinguishable
Bose gases \cite{BCKR,CBK} which interact via an effectively-repulsive 
contact potential, and are confined in a two-dimensional concentric 
double-ring-like trap, as shown in Fig.\,\ref{fig1}. 
Using the mean-field approximation, we investigate two main 
questions: First, we identify the various phases in the ground 
state of the system, varying the interaction strength between 
the atoms. In the trapping potential that we consider, we 
observe that the two gases separate radially via discontinuous 
transitions; in this case, each gas resides in one of the two 
minima of the trapping potential, preserving the circular 
symmetry of the trapping potential. We also observe the expected 
azimuthal (and continuous) phase separation between the two 
gases in each potential minimum \cite{Ao,Timm,PS}. 
A similar effect has also been studied in the case of a 
single ring \cite{Smy09}; see also \cite{Ao,Timm,PS}.

The second main question that we examine are the rotational
properties of this system, including its response to some
rotational frequency of the trap $\Omega$, as well as the
stability of the persistent currents for variable couplings,
and variable relative populations of the two components.
The expectation value of the angular momentum of the
system as a function of $\Omega$ shows an interesting
structure, reflecting the various phase transitions that
take place with increasing $\Omega$.

Regarding the (meta)stability of the currents, it is remarkable
that for equal populations between the two components the vast 
majority of the coupling strengths that we have examined yield 
metastable states, except for a very small range where all the 
coupling strengths are exactly or nearly equal.

It is worth mentioning that analogous single and concentric ring
geometries have been addressed in semiconductor heterostructures,
both theoretically and experimentally, see e.g. \cite{CQRs}, and
also \cite{reviews} for reviews on the subject. In these systems, 
the applied external magnetic field plays the same role as the 
trap rotation in the present problem and allows the investigation
of, e.g., electron localization effects and persistent electron
currents in field-free regions.

In what follows we first describe in Sec.\,II our model. In
Sec.\,III we present the results for the ground state of the
system, identifying the states where the species coexist, or
separate, either radially or azimuthally. In Sec.\,IV we examine
the rotational properties for a fixed rotational frequency of
the trap, and the (meta)stability of the persistent currents. 
We study the stability as a function of the coupling between the 
atoms, as well as of the ratio of the populations of the two 
components. Finally, in Sec.\,V we present a summary and our 
conclusions.

\section{Model and method}

\begin{figure}[t]
\centerline{\includegraphics[width=7cm,clip]{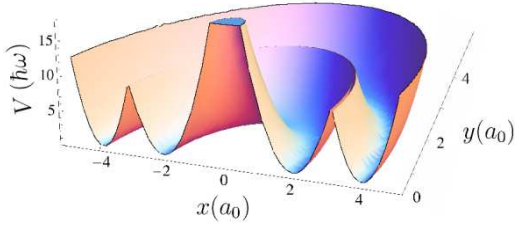}}
\caption{(Color online)
The confining potential $V$ of Eq.\,(\ref{trpot}), where $R_{i}=2 a_0$, 
$R_{o} = 4 a_0$, and $\omega_{o}/\omega_{i} = 5/4$.}
\label{fig1}
\end{figure}

We consider two distinguishable kinds of bosonic atoms, labelled as
$A$ and $B$, which are trapped in a two-dimensional potential
of the form
\begin{equation}
V(\rho) =  {\rm min} \left\{
\frac{1}{2}\, M \omega_{i}^2\, (\rho - R_{i})^2 \; , \;
\frac{1}{2}\, M \omega_{o}^2\, (\rho - R_{o})^2 \right\},
\label{trpot}
\end{equation}
where $\rho$ is the usual radial coordinate in cylindrical
coordinates and $M$ is the atom mass, assumed to be equal for
the two components. The two (overlapping) parabolae in $V(\rho)$
with frequencies $\omega_i$ and $\omega_o$ are centered at
the positions with $\rho = R_i$ and $\rho = R_o$, giving rise
to the potential plotted in Fig.\,\ref{fig1} \cite{confinement}.

In our calculations we consider $R_{i}=2 a_0$ and $R_{o}=4
a_0$, where $a_0 = [\hbar/(M \omega)]^{1/2}$ is the oscillator
length corresponding to $\omega = \omega_{i}/4$, and finally
$\omega_{o}/\omega_{i} = 5/4$. In the outer ring the potential
is more tight, $\omega_{o} > \omega_{i}$, in order to compensate
for the fact that $R_{o} > R_{i}$, i.e., to make the product of
the ``width" of each annulus times the radius of each annulus
to be comparable to each other. (This is typically also the case
in the studies on electrons in quantum rings, in semiconductor
heterostructures that we mentioned above.) 

To simplify the discussion, we also assume that there is a very 
tight trapping potential along the $z$ axis (omitted in the potential 
above), which completely freezes out the degrees of freedom of the 
gases along this direction. With this assumption, our problem 
becomes effectively two-dimensional, with the tight dimension 
entering only implicitly through the parameters $u_{ij}$ in the 
Hamiltonian of Eq.\,(\ref{Ham}). With the usual assumption of a 
contact interatomic potential, the Hamiltonian becomes
\begin{eqnarray}
   H = \sum_{i=1}^{N_A} - \frac {\hbar^2} {2 M}
\nabla_{i}^2 + V(\rho_i) +
\sum_{j=1}^{N_B} - \frac {\hbar^2} {2 M}
\nabla_{j}^2 + V(\rho_j) +
\nonumber \\
+ \frac {u_{AA}} 2 \sum_{i \neq j=1}^{N_A}
\delta({\bf r}_{i} - {\bf r}_{j})
+ \frac {u_{BB}} 2 \sum_{i \neq j=1}^{N_B}
\delta({\bf r}_{i} - {\bf r}_{j})
\nonumber \\
+ u_{AB} \sum_{i=1, j=1}^{N_A, N_B}
\delta({\bf r}_{i} - {\bf r}_{j}).
\label{Ham}
\end{eqnarray}
Here $u_{ij} = (4 \pi \hbar^2 a_{ij}/M) \int |\psi_0(z)|^4 \, dz$, 
with $\psi_0(z)$ being the state of lowest energy of the potential
along the $z$ axis and $a_{ij}$ being the $s-$wave scattering lengths 
for zero-energy elastic atom-atom collisions. The coupled 
Gross-Pitaevskii-like equations for the order parameters of the 
two components $\phi_A$ and $\phi_B$, resulting from the above 
Hamiltonian, are
\begin{eqnarray}
- \frac {\hbar^2 \nabla^2}{2 M} {\phi}_A + V(\rho) {\phi}_A
+ g_{AA} |{\phi}_A|^2 {\phi}_A + g_{AB} |{\phi}_B|^2 {\phi}_A
\nonumber \\ = {\mu}_A {\phi}_A,
\nonumber \\
- \frac {\hbar^2 \nabla^2}{2 M} {\phi}_B + V(\rho) {\phi}_B
+ g_{BB} |{\phi}_B|^2 {\phi}_B + g_{AB} |{\phi}_A|^2 {\phi}_B
\nonumber \\
= {\mu}_B {\phi}_B.
\label{gpe}
\end{eqnarray}
In the above equations $\int |{\phi}_A|^2 d {\bf r} = N_A/N_B$,
and $\int |{\phi}_B|^2 d {\bf r} = 1$. Also, $g_{AA} =
N_B u_{AA}$, $g_{BB} = N_B u_{BB}$, $g_{AB} = N_B u_{AB}$,
and $\mu_i$ are the chemical potentials. In what follows,
we consider repulsive interactions only, $g_{ii} > 0$,
$g_{ij}>0$, and also assume that $g_{AA} = g_{BB} \equiv g$.

The method that we adopt to solve Eqs.\,(\ref{gpe}) is a
fourth-order split-step Fourier method within an imaginary-time
propagation approach \cite{Chi05}. We start with a reasonable
initial state for the two components and propagate it in imaginary 
time, making sure that we proceed a sufficiently large number of 
time steps, which guarantee that we have reached a steady state.

\section{Phase diagram of the ground state}

We start with the ground state of the system. There are
three energy scales in the problem, namely the single-particle
energy that is set by the trap, the intra-atomic interaction
energy, and the inter-atomic interaction energy. For weak
interactions, the energy is dominated by the single-particle
term and the ground state is determined by the minimization
of this term. On the other hand, for strong interactions, for 
a given ratio of the two populations $N_A/N_B$, the actual 
symmetry of the ground state results from the competition 
between the intra-and inter-species coupling strengths; the 
former favors the maximum possible spread of the two gases 
within the system, whereas the latter favors the minimization 
of their spatial overlap.

The most pronounced difference of this problem as compared
to the case where there is only one potential minimum -- i.e.,
a single annulus -- is the existence of a phase where each component
resides in only one of the two potential minima, thus separating 
radially. In addition, we have also observed the expected azimuthal
phase separation when both species occupy the same potential
minimum \cite{Smy09,Ao,Timm,PS}.

\begin{figure}[t]
\centerline{\includegraphics[width=8cm,clip]{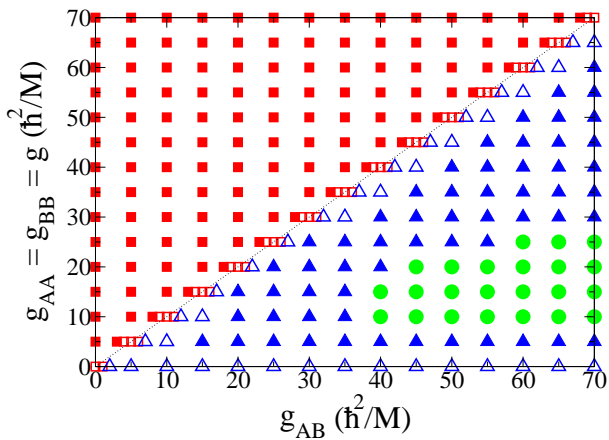}}
\caption{(Color online)
Phase diagram in the $g_{AB}$-versus-$g$ plane, which
shows the symmetry of the ground-state density distribution of
the two components. Squares, triangles and circles correspond to
phase coexistence, azimuthal phase separation, and radial phase
separation, respectively. The solid (empty) symbols denote points
where the currents are metastable (unstable).}
\label{fig2}
\end{figure}

In Fig.\,\ref{fig2} we illustrate the phase diagram showing the
symmetry of the ground-state density distribution of the two gases
as $g_{AB}$ and $g$ are varied, for $N_A = N_B$. Since $g_{AA} =
g_{BB}$ and $N_A = N_B$, the results are symmetric when the two
components are interchanged. The axes in the phase diagram of
Fig.\,\ref{fig2} may also be considered to represent the values
of the scattering lengths $a_{AB}$ and $a_{AA} = a_{BB}$ (scaled
appropriately).

\begin{figure}[t]
\centerline{\includegraphics[width=9cm,clip]{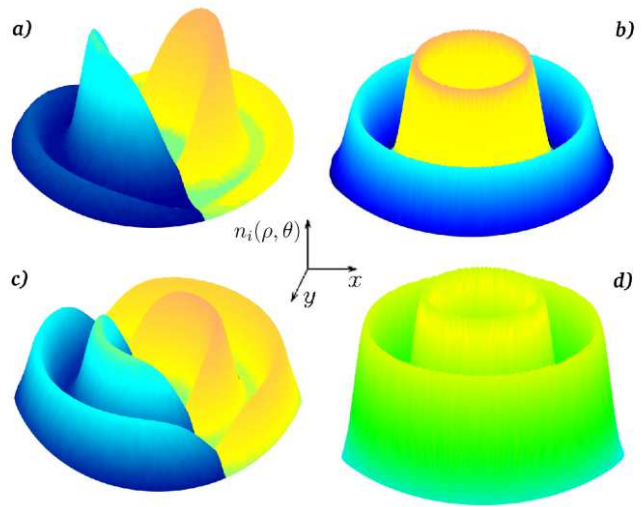}}
\caption{(Color online)
Ground-state density distributions $n_A(\rho,\theta) = 
|\phi_A(\rho,\theta)|^2$ and $n_B(\rho,\theta) = |\phi_B(\rho,\theta)|^2$ 
of the two components, for $g_{AB} = 55$ and (a) $g = 5$, (b) $g = 15$, 
(c) $g = 35$, and (d) $g = 60$, corresponding to azimuthal phase 
separation, radial phase separation, azimuthal phase separation, and 
phase coexistence, respectively. The peaks in the densities correspond 
to the minima of the inner and the outer parabolae of the confining 
potential.}
\label{fig3}
\end{figure}

We have found three different phases: (i) coexistence of the two 
species (squares, red color), (ii) azimuthal separation (triangles, 
blue color), and (iii) radial separation (circles, green color), see 
Fig.\,\ref{fig3}. The difference between solid and empty symbols in 
Fig.\,\ref{fig2} refers to the stability of the persistent currents 
and is explained in the following section.

As seen in Fig.\,\ref{fig2}, the phase boundary between phase
coexistence and separation of the two components is, to a very
good approximation, a straight line given by (setting $\hbar = M
= \omega = 1$)
\begin{equation}
 g_{AB} \approx g + 2,
\label{ps}
\end{equation}
as we have found by fitting numerically our data.

Certain limiting cases in the phase diagram of Fig.\,2 may be
analyzed and understood easily. In the case $g_{AB} = 0$
and $g_{AA} = g_{BB}$, the two gases do not interact with each
other. In order to minimize their energy, they distribute
homogeneously along the rings and coexist. In the other
limiting case $g_{AA} = g_{BB} = 0$, for $g_{AB}$ smaller than
$\approx 2$, the two components also coexist; however, for larger 
values than 2, they  separate azimuthally. Although in this phase
the kinetic energy increases, the interaction energy is lowered
due to the repulsion between the two species, and the
azimuthal symmetry-breaking persists with increasing $g_{AB}$.

It is also instructive to understand the internal structure
of the phase diagram. As one moves vertically, i.e.,
for a fixed value of $g_{AB}$ (being sufficiently large,
such that the components separate azimuthally), for small
enough $g$ the two components occupy mainly the inner ring since
the repulsion between the particles cannot compensate for
the stronger confinement of the outer ring. As $g$
increases, the outer ring becomes progressively more
occupied. When the inter-component repulsion
$g_{AB}$ becomes large enough, the gases minimize their energy
by separating radially. However, if $g$ increases further,
the inner ring becomes too small to host one of the species
entirely. Azimuthal phase separation takes place again, now
with both gases being largely spread within the whole system.
Eventually, for even larger values of $g$, the dominant term
in the energy is the intra-atomic interaction, and thus the two
components coexist.

As anticipated before, we illustrate this effect in Fig.\,\ref{fig3},
where we plot the densities of the cases with $g_{AB} = 55$, 
and (a) $g = 5$, (b) $g = 15$, (c) $g = 35$, and (d) $g = 60$,
corresponding to azimuthal phase separation, radial phase
separation, azimuthal phase separation, and phase coexistence,
respectively.

It is also of interest to investigate the nature of the phase
transitions occurring in the system. As one crosses the boundary from
coexistence to azimuthal phase separation, the two components decrease
continuously their overlap, developing sharper profiles as the repulsion
increases. This transition is thus continuous (second order), as it is
also the case in purely one-dimensional single rings \cite{Smy09}. In the
corresponding energy surface, the minimum (which determines the ground 
state) moves continuously as one crosses the phase boundary. On the 
contrary, the transitions involving radial separation are discontinuous 
(first order), indicating that two local minima in the energy surface 
compete and that the system jumps abruptly from the one state to the other.

\section{Rotational properties}

\subsection{Lowest state of the system for a fixed angular
frequency of rotation}

\begin{figure}[t]
\centerline{\includegraphics[width=9cm,clip]{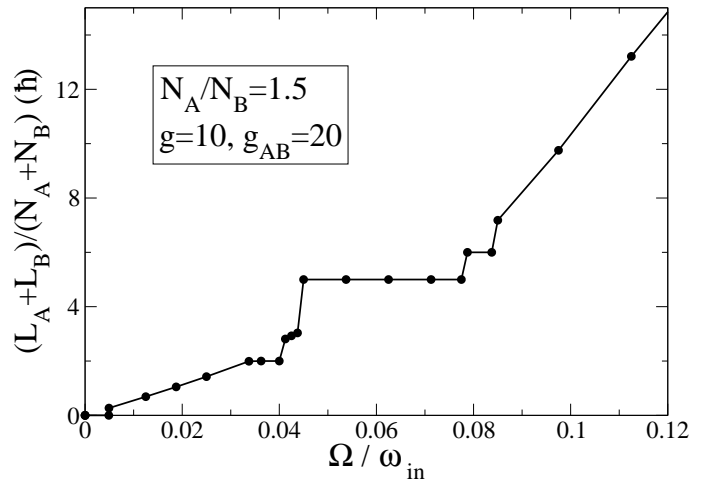}}
\caption{(Color online)
The total angular momentum per particle, $(L_A+L_B)/(N_A + N_B)$,
as a function of the angular frequency of rotation of the trap
$\Omega$, which results from the minimization of the energy in
the rotating frame.}
\label{fig4}
\end{figure}

The phase transitions that we described above also have
a clear influence on the rotational properties of the system.
We start by examining the response of the system to some finite
rotational frequency $\Omega $ of the trap. In the following
we determine the total angular momentum per particle $l =
(L_A + L_B)/(N_A + N_B)$ as a function of $\Omega$, $l =
l(\Omega)$. Following the usual procedure, we minimize the
energy of the system in the rotating frame, i.e., we minimize
$E/N - l \Omega$, where $E$ is the total energy. The result
of this calculation is shown in Fig.\,\ref{fig4}, where we
have set $N_A/N_B=1.5$, $g = 10$, and $g_{AB} = 20$. The
typical values of $\Omega$ are rather small because we have
scaled it with $\omega_i$. However, at least in the
case of a purely one-dimensional ring potential, the scale
for the typical $\Omega$ is on the order of the frequency
$\hbar/(2 M R^2)$ of the corresponding kinetic energy
$\hbar^2/(2 M R^2)$ \cite{Leggett}, where $R$ is the
radius of the ring. In atomic units, $\hbar = M = \omega
= 1$, $\omega_i = 4$, while $\hbar/(2 M R^2) \simeq 1/50$
(for $R$, say, equal to 5). This introduces a factor
$10^{-2}$.

It is also instructive to comment on the behavior of the gas
as $\Omega$ increases and see the connection of the function
$l = l(\Omega)$ with the density distribution of the two
components. For $0 \le \Omega/\omega_i \le 0.005$ the two
species are separated radially and the total angular
momentum is zero. For $\Omega/\omega_i \approx 0.005$, the
density of the two components breaks its azimuthal symmetry
discontinuously, and the angular momentum jumps abruptly to 
a finite value. Beyond this value of $\Omega/\omega_i$
the angular momentum increases linearly with $\Omega$ and 
is carried by both components, which undergo solid-body rotation. 
This is an expected result due to the azimuthal symmetry-breaking 
of the density, as we have confirmed by studying the phases of 
the two order parameters. When $\Omega/\omega_i \approx 0.035$, 
a new plateau appears, which corresponds to an angular momentum 
per particle equal to two, and the two species separate radially.
As one can see from the plot, there is a sequence between
plateaus at integer values of $l$ and straight lines
with a positive slope, which are separated by abrupt jumps.
This sequence persists up to $\Omega/\omega_i \approx 0.085$.
Beyond this value of $\Omega$ there is no longer radial separation
of the two components due to the large centrifugal force, which 
forces both gases to occupy the two potential minima and therefore 
to separate azimuthally. 

Similar discontinuous transitions in the function $l = l(\Omega)$
occur in a single-component weakly-interacting Bose-Einstein
condensate that rotates in a harmonic trap \cite{Rokhsar}, and 
are associated with discontinuous transitions between phases of 
different symmetries of the single-particle density distribution 
of the gas. In the present problem, the corresponding different
symmetries are the ones where the components separate radially,
or azimuthally.

Furthermore, the jumps in the $l = l(\Omega)$ plot are
consistent with the fact that for the parameters considered,
the system supports persistent currents. In other words,
had this function been continuous, then metastability would
have not been possible (for a discussion of this effect
we refer to Ref.\,\cite{Leggett}).

From the above observations we see that the general picture 
that emerges by considering a finite rotation of the trap 
resembles the one where the couplings are varied, with a 
series of phase transitions between radial and azimuthal 
separation.

\subsection{Metastable currents}

Another interesting question is the possible existence of 
(meta)stable currents. In physical terms, we investigate the 
energetic (meta)stability of current-carrying states. In the 
case that there is an energy barrier that separates a 
current-carrying state from the ground state, in the presence 
of some dissipative mechanism (such as a thermal cloud, for 
example) such a state does not decay, and therefore the system
supports persistent currents.

We examine two separate aspects of this problem. Firstly, we
consider the points of the phase diagram shown in Fig.\,\ref{fig2},
for a fixed population of the two gases. Secondly, we fix the
interaction strengths and vary the ratio $N_A/N_B$. In both cases
we examine the imaginary-time evolution of initial states with some 
finite, nonzero expectation value of the angular momentum, in the 
absence of any external rotation of the trap. For a given set of 
parameters, the existence of a converged final state with a nonzero 
expectation value of $l$ implies that the associated current is 
metastable.

\subsubsection{Variable couplings and fixed populations}

We have examined all the points that are shown in the phase
diagram of Fig.\,\ref{fig2}. Those corresponding to a final
state with a nonzero expectation value of the angular momentum
are represented in Fig.\,2 with ``solid'' symbols, while the
ones that decay to the non-rotating ground state are represented 
with ``open'' symbols. 

Clearly, the vast majority of states correspond to metastable 
currents, except those close to the diagonal $g = g_{AB}$, as 
well as those with sufficiently small $g$ (for all values of 
$g_{AB}$). The obtained results show that the angular momentum 
of the metastable states is always an integer multiple of the 
particle population, which implies that the associated densities 
are necessarily circularly symmetric. This is consistent with the 
statement that circular symmetry is a necessary -- though not 
sufficient -- condition for the (meta)stability of the currents, 
as otherwise the circulation may escape from the gas (since in 
this case there is no barrier separating the rotating state from 
the non-rotating one \cite{Leggett}).

\subsubsection{Variable relative population and fixed couplings}

Let us now study the effect of a variable relative population
between the two gases on the (meta)stability of the currents. We
thus fix the interaction strengths, as well as the population of
the one component $(N_B)$, and study this question starting from the
case with $N_A=0$, all the way up to the limit $N_A \gg N_B$, with
an initial state that has some angular momentum in component $B$.
Since $N_B$ is fixed, the above procedure corresponds physically to
keeping the scattering lengths fixed.

As mentioned above, the (meta)stability of the persistent currents 
depends on the competition between azimuthal phase separation and 
circular symmetry of the density distribution of the two components. 
Thus, when $N_A=0$, the $B$ component spreads within the whole system 
for any value of $g$ and the currents can be metastable, provided that 
the coupling $g$ is sufficiently large. If the population of the component 
$A$ becomes nonzero but is small enough, it acts only as a weak 
perturbation, and azimuthal symmetry is still preserved.

However, beyond a critical ratio $N_A/N_B$, the two species separate
azimuthally and the currents are no longer metastable. This critical
value depends on the actual intra- and inter-component interaction
strengths. A further increase of $N_A$ with respect to $N_B$
drives the system to a phase of radial separation, and metastability 
is recovered.

Finally, in the limit where $N_A$ becomes too large, this component
cannot fit into only one of the two rings. This leads again to azimuthal
phase separation of the gases, which is preserved even in the limit
$N_A \gg N_B$. As a result, the stability of the currents is lost
again.

\begin{figure}[t]
\centerline{\includegraphics[width=8cm,clip]{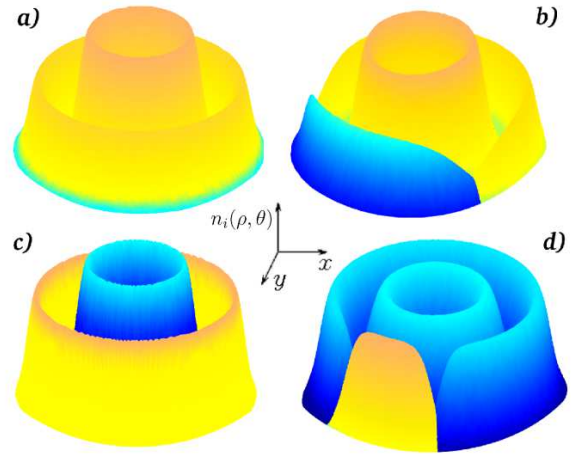}}
\caption{(Color online)
Density distributions $n_A(\rho,\theta) = |\phi_A(\rho,\theta)|^2$ and 
$n_B(\rho,\theta) = |\phi_B(\rho,\theta)|^2$ with $g = 35$, and $g_{AB}=55$ 
of the component $A$ (dark blue) and $B$ (light yellow), for 
(a) $N_A/N_B = 0.01$, (b) $N_A/N_B = 0.25$, (c) $N_A/N_B = 0.75$, 
and (d) $N_A/N_B = 10$. In the cases (a) and (c) the states support 
metastable currents, while in (b) and (d) they do not. The peaks 
in the densities correspond to the minima of the inner and the outer 
parabolae of the confining potential.}
\label{fig5}
\end{figure}

We show in Fig.\,\ref{fig5} the densities for the case $g = 35$,
and $g_{AB} = 55$ and for various values of the ratio $N_A/N_B$,
in order to illustrate the mentioned sequence of phase transitions.
In particular, the results correspond to $N_A/N_B = 0.01$, $N_A/N_B
= 0.25$, $N_A/N_B = 0.75$, and $N_A/N_B = 10$, with the first and
the third cases corresponding to the metastable states. According 
to our simulations, changing the values of the interaction strengths 
modifies the sizes of the ``windows'' in $N_A$ that separate the 
different phases, but yields qualitatively similar results.

\section{Summary and conclusions}

As shown in the present study, a coupled system of two distinguishable
Bose gases that interact with an effectively-repulsive contact 
potential and are loaded in a concentric double annular trap reveals 
a series of phase transitions and the existence of metastable 
currents.

For weak interactions, when the chemical potential is much
smaller than the barrier that separates the two potential
minima, the gases are confined in the inner ring, with the
width of their transverse profile being smaller than
the radius of the ring. Thus, their motion is (at least)
close to being quasi-one-dimensional. On the other hand,
as the couplings increase, the barrier plays a decreasingly
important role, and their transverse width becomes comparable
to the radius of the ring(s). Such a trapping potential 
interpolates up to some extent between one-dimensional
and two-dimensional motion, depending on the strength of
the coupling between the atoms. The interplay between this
effect and the strength of the inter- and intra-atomic
couplings gives rise to interesting phase transitions
in the ground state of the system, including three different 
geometries: phase coexistence, radial phase separation, and 
azimuthal phase separation.

An interesting feature of the system considered is its response 
to some finite rotational frequency of the trap. The basic picture 
resembles very much the one where the couplings are varied, 
inducing axial and/or radial phase separation of the two components.

The robustness in the (meta)stability of the currents that we
found for the vast majority of the points in the phase diagram
of Fig.\,\ref{fig2} is another interesting aspect of this study.
Metastability is not found for the cases when the couplings are 
all nearly equal or exactly equal to each other, as well as for 
the cases with small $g$, independently of $g_{AB}$.

According to Ref.\,\cite{metastability}, a necessary (but not
sufficient) condition for metastability is that the trapping 
potential does not increase monotonically from the center
of the trap. The present results suggest that multiple
variations in the monotonicity of the trapping potential enhance
the stability of the currents.

Last but not least, the discontinuous phase transitions we have
found both in the ground-state phase diagram when the two gases
separate radially as the couplings are varied, as well as in the 
response of the system when the rotation of the trap is varied, 
imply that hysteresis should show up as the coupling/rotational 
frequency increases/decreases.

\section{Acknowledgements}

We thank S. Bargi and K. K\"arkk\"ainen for useful discussions. This work
was financed by the Swedish Research Council. The collaboration is part of
the NordForsk Nordic network ``Coherent Quantum Gases - From Cold Atoms
to Condensed Matter''.


\begin{thebibliography}{99}

\bibitem{Gupta} S. Gupta, K. W. Murch, K. L. Moore, T. P. Purdy,
and D. M. Stamper-Kurn, Phys. Rev. Lett. {\bf 95}, 143201 (2005).

\bibitem{Olson} S. E. Olson, M. L. Terraciano, M. Bashkansky,
and F. K. Fatemi, Phys. Rev. A {\bf 76}, 061404(R) (2007).

\bibitem{Ryu} C. Ryu, M. F. Andersen, P. Clad\' e, V. Natarajan,
K. Helmerson, and W. D. Phillips, Phys. Rev. Lett. {\bf 99}, 260401
(2007).

\bibitem{Hen} K. Henderson, C. Ryu, C. MacCormick, and M. G. Boshier,
New J. Phys. {\bf 11}, 043030 (2009).

\bibitem{vK} Igor Lesanovsky and Wolf von Klitzing, Phys. Rev. Lett.
{\bf 99}, 083001 (2007).

\bibitem{Wolf} Igor Lesanovsky and Wolf von Klitzing, Phys. Rev. Lett.
{\bf 98}, 050401 (2007).

\bibitem{Brand} J. Brand, T. J. Haigh, and U. Z\"ulicke, Phys. Rev.
A {\bf 80}, 011602(R) (2009).

\bibitem{Smy09} J. Smyrnakis, S. Bargi, G. M. Kavoulakis,
M. Magiropoulos, K. K\"arkk\"ainen, and S. M. Reimann Phys.
Rev. Lett. {\bf 103}, 100404 (2009).

\bibitem{BCKR} S. Bargi, J. Christensson, G. M. Kavoulakis,
and S. M. Reimann, Phys. Rev. Lett. {\bf 98}, 130403 (2007).

\bibitem{CBK} J. Christensson, S. Bargi, K. Karkkainen, Y. Yu,
G. M. Kavoulakis, M. Manninen, S. M. Reimann, New J. Phys.
{\bf 10}, 033029 (2008).

%Building such traps is possible using various
%techniques (Wolf von Klitzing, private communication).

\bibitem{Ao} P. Ao and S. T. Chui, Phys. Rev. A {\bf 58}, 4836
(1998).

\bibitem{Timm} E. Timmermans, Phys. Rev. Lett. {\bf 81}, 5718
(1998).

\bibitem{PS} C. J. Pethick and H. Smith, {\it Bose-Einstein
Condensation in Dilute Gases} (Cambridge University Press,
Cambridge, England, 2002).

\bibitem{CQRs} B. Szafran and F. M. Peeters, Phys. Rev. B {\bf 72},
155316 (2005); J. M. Escart\'in, F. Malet, A. Emperador, and M. Pi,
Phys. Rev. B {\bf 79}, 245317 (2009); T. Mano, T. Kuroda,
S. Sanguinetti, T. Ochiai, T. Tateno, J. Kim, T. Noda,
M. Kawabe, K. Sakoda, G. Kido, and N. Koguchi, Nanoletters {\bf 5}, 425
(2005); S. Viefers, P. S. Deo, S. M. Reimann, M. Manninen, and
M. Koskinen, Phys. Rev. B {\bf 62}, 10668 (2000); 
M. Manninen, M. Koskinen, S. M. Reimann, and B. Mottelson,
Eur. Phys. J. D {\bf 16}, 381.

\bibitem{reviews} S. Viefers, P. Koskinen, P. S. Deo,
and M. Manninen, Physica E {\bf 21}, 1 (2004);
S. M. Reimann and M. Manninen, Rev. Mod. Phys. {\bf 74}, 1283 (2002).  

\bibitem{confinement} Similar trap geometries have been considered
in coupled quantum rings, see e.g. the first and second papers of
\cite{CQRs}.

\bibitem{Chi05} S. A. Chin and E. Krotscheck, Phys. Rev. E {\bf 72},
036705 (2005).

\bibitem{Leggett} A. J. Leggett, Rev. Mod. Phys. {\bf 73},
307 (2001).

\bibitem{Rokhsar} D. A. Butts, and D. S. Rokhsar, Nature (London)
{\bf 397}, 327 (1999).

\bibitem{metastability} K. K\"arkk\"ainen, J. Christensson, G. Reinisch,
G. M. Kavoulakis, and S. M. Reimann, Phys. Rev. A {\bf 76}, 043627
(2007).

\end{thebibliography}
\end{document}